\documentclass[12pt,preprint]{aastex}

\usepackage{amsmath}
\usepackage{amssymb}

\shorttitle{Thermodynamics of partially relaxed stellar systems}
\shortauthors{Bertin and Trenti}

\begin{document}

\title{Thermodynamical description of a family of partially relaxed stellar systems}

\author{G. Bertin}
\affil{Dipartimento di Fisica, Universit\`{a} di Milano, via
Celoria 16, I-20133 Milano, Italy}
\email{Giuseppe.Bertin@unimi.it}

\and

\author{M. Trenti}
\affil{Scuola Normale Superiore, piazza dei Cavalieri 7, I-56126 Pisa, Italy}
\email{m.trenti@sns.it}

\begin{abstract}
We examine the thermodynamical
properties of a family of partially relaxed, anisotropic stellar
systems, derived earlier from the Boltzmann entropy under the
assumption that a third quantity $Q$ is conserved in addition to
the total energy and the total number of stars.  We now show that the family
of models conforms to the paradigm of the gravothermal
catastrophe, which is expected to occur (in the presence of
adequate energy transport mechanisms) when the one-parameter
equilibrium sequence attains sufficiently high values of the
concentration parameter; these are the values for which the models
are well fitted by the $R^{1/4}$ law. In the intermediate
concentration regime the models belonging to the sequence exhibit
significant deviations from the $R^{1/4}$ law. Curiously, in the
low-concentration regime, the global thermodynamical temperature
associated with the models becomes negative when the models become
too anisotropic so that they are unstable against the radial orbit
instability; this latter behavior, while offering a new clue to
the physical interpretation of the radial orbit instability, is at
variance with respect to the low-concentration limit of the
classical case of the isotropic, isothermal sphere investigated by
\citet{bon56}  and \citet{lyn68}.
\end{abstract}


\section{Introduction}
The possibility of providing a thermodynamical description of
self-gravitating stellar systems has motivated a number of
investigations in galactic dynamics, starting with the pioneering work
of Antonov and Lynden-Bell in the 60s. After the realization that
violent relaxation is likely to lead to partially relaxed
configurations in dynamical equilibrium \citep{lyn67}, a
re-examination of the problem of the isothermal sphere, studied
earlier by \citet{bon56} for a self-gravitating gas, led to the
interesting possibility that stellar systems may undergo the process
of gravothermal catastrophe (\citealt{lyn68}; see also
\citealt{ant62}).  However, a rigorous derivation of the onset of the
gravothermal catastrophe from a study of the Boltzmann entropy
appeared to be available only for the case of ideal systems confined
by a spherical reflecting wall. A number of convincing qualitative
arguments made it clear that also unbound stellar systems with finite
mass, such as those described by the King sequence (\citealt{kin66};
these spherical models have a finite radius, but do not require an
external wall), should probably fall into the same physical framework
and indeed the paradigm received a lot of attention, especially in the
context of the dynamics of globular clusters (see \citealt{spi87}),
which are known to possess, at least to some extent, the desired
internal collisionality (see also \citealt{lyn80}).  [An indirect
indication that the general physical picture of the gravothermal
catastrophe is likely to be robust comes also from the proof that the
behaviour of the classical gas case is basically independent of the
assumption of spherical geometry \citep{lom01}.]  Several
investigations have aimed at producing a rigorous derivation of the
gravothermal catastrophe for unbound stellar systems, focusing on the
underlying argument that refers to the Poincar\'{e} stability of
linear series of equilibria \citep{kat78,kat79,pad89}, but the proof
has always been centered on an unjustified {\it ansatz} in order to
connect the underlying entropy $S$ with the global temperature $T =
1/(\partial{S}/\partial{E_{tot}})$ (see Appendix V in the article by
\citealt{lyn68}; \citealt{kat80}; \citealt{mag98}). Other
investigations have explored the possibility of setting the discussion
in the context of non-standard entropies (e.g., the Tsallis entropies;
see the study of the polytropic spheres by \citealt{cha02} and
\citealt{tar02}). Note that the concept of entropy for collisionless
systems is quite subtle (e.g., see \citealt{sti87} and references
therein). One might even argue whether it is actually compatible with
the long-range nature of gravity, given the fact that self-gravitating
systems lack additivity, a key ingredient in thermodynamics.

In the meantime, inspired by N-body simulations of collisionless
collapse \citep{van82}, which confirmed the general picture of
incomplete violent relaxation and showed that it can lead to systems
with realistic density profiles without ad hoc tuning of the initial
conditions, some families of models were constructed able to
reproduce, for quasi-spherical configurations, the characteristic
feature of the anisotropy profile with an inner isotropic core and an
outer radially biased envelope (\citealt{ber84}; see \citealt{ber93}
and references therein): these families turned out to exhibit the
characteristic $R^{1/4}$ projected density profile and indeed were
shown to match nicely the observed photometric and kinematical
characteristics of bright ellipticals. In an attempt at providing a
justification of these models (in particular, of the so-called
$f_{\infty}$ models, constructed initially only from dynamical
arguments) from statistical mechanics, two routes were pursued
\citep{sti87}.

The first combines an
explicit statement of partial relaxation, i.e. of a relaxation process
that is expected to be inefficient in the outer regions, and
the existence of a suitable weight, related to the orbital period,
for the cells that make the relevant partition of phase space; it follows
qualitative arguments proposed by \citet{lyn67} and is  physically appealing (see also \citealt{tre86a}).
It was indeed shown to lead naturally to the $f_{\infty}$ models.
However, this route is not fully satisfactory from the
mathematical point of view, especially since it involves an
explicit approximation for the orbital period that is applicable
only to the low binding energy limit of quasi-Keplerian orbits.
The second route is straightforward from the mathematical point of
view, being based on the classical Boltzmann entropy and on the
assumed explicit conservation of a third quantity $Q$, in
addition to the total mass $M$ and to the total energy $E_{tot}$.
It was shown to lead to an analytically different family of models
(that we may call the $f^{(\nu)}$ models; see definition in
Sect.~3 below), with qualitative properties very similar to those
of the $f_{\infty}$ models. Those models were not studied much
further and did not receive great attention, not only because the
relevant distribution function is not as simple as that of the
$f_{\infty}$ models, but especially because the conservation of
$Q$ could only be justified approximately by inspection of a
number of N-body simulations, without a clear-cut physical
justification (see \citealt{sti87}; in contrast, the
conservation of the additional ${\bf A} \cdot {\bf B}$ invariant
sometimes invoked in plasma physics is rather transparent; see
\citealt{cha58}).

In this paper we take advantage of the simple statistical
mechanics foundation of the $f^{(\nu)}$ family of unbound
partially relaxed stellar systems to explore the possibility of a
thermodynamical description of stellar-dynamical models that are
endowed with realistic properties.

\section{Comment on the evolution of elliptical galaxies}

Before proceeding to illustrate the results of this paper, we make a short digression in order to bring out the connections between the present analysis and the evolution of elliptical galaxies. We start by recalling that, formally, 
the sequence of \citet{kin66} models is one special family of solutions of
the collisionless Boltzmann equation. Yet, it is recognized to provide a
reasonable description of the current properties of globular clusters
(see \citealt{djo94}), within a framework where these stellar
systems continually evolve as a result of a variety of mechanisms (among
which star evaporation and disk shocking; see \citealt{ves97} and
references therein) and where the paradigm of the gravothermal
catastrophe can be applied (see \citealt{spi87}). Of course, it is well
known that the level of internal collisionality in globular clusters is
relatively high, so that the above approach is quite natural.

In contrast, one might at first think of dismissing the possibility
that the paradigm of the gravothermal catastrophe should be of
interest for the study of elliptical galaxies, because these large
stellar systems lack the desired level of collisionality, judging from
the estimate of the relevant star-star relaxation times. Here,
following the spirit of earlier investigations (starting with
\citealt{lyn68}), we note that real elliptical galaxies are actually
complex systems the evolution of which goes well beyond the idealized
framework of the collisionless Boltzmann equation. In other words,
splitting their description into past (formation) processes and
present (mostly collisionless equilibrium) conditions should be
considered only as an idealization introduced in order to assess the
properties that define their current basic state. 

In practice, elliptical galaxies are expected to be in a state of
continuous evolution, for which we can list several specific dynamical
causes: (1) Left-over granularity of the stellar system itself from
initial collapse.  Clumps of stars are likely to continue to form and
dissolve in phase space even after the system has reached an
approximate steady state.  This acts as internal collisionality thus
making some relaxation proceed even at current epochs. Indeed,
numerical simulations of violent (partial) relaxation show that some
evolution continues well after the initial collapse has taken
place. (2) Drag of a system of globular clusters or other heavier
objects towards the galaxy center. A globular cluster system or the
frequent capture of small satellites (mini-mergers) may provide an
internal heating mechanism associated with the process of dynamical
friction by the stars on the heavier objects \citep{ber02b}. (3)
Long-term action of tidal interactions of the galaxy with external
objects. (4) Presence of gas in various phases (cold, warm, and
hot). Significant cooling flows have been observed in bright
ellipticals. Traditionally, studies of processes of this kind focus on
the dynamics of the cooling gas and keep the background stellar system
as `frozen'. In reality, energy and mass exchanges take place between
the stellar system and the interstellar medium. (5) Interaction
between the galactic nucleus and the galaxy. A number of interesting
correlations have been found between the properties of galaxy nuclei
and global properties of the hosting galaxies (e.g., see
\citealt{pel99}). These correlations suggest that significant energy
exchanges are taking place between the galaxy and its
nucleus. Eventually, if a sufficiently concentrated nucleus is
generated, then star-star relaxation in the central regions may also
become a significant cause of dynamical evolution.

All of the above are specific mechanisms that are expected to make
elliptical galaxies evolve in spite of their very long typical
star-star relaxation time. Most of these processes are hard to model
and to calculate in detail. As for the evolution of other complex
many-body systems, it is hoped that thermodynamical arguments may help
us identify general trends characterizing such evolution. This is the
basic physical scenario in which the calculations presented in this
paper are expected to be of interest for real elliptical
galaxies.

\section{Partially relaxed, unbound, finite mass systems from the Boltzmann entropy}

Let us consider the standard Boltzmann entropy $S = - \int f \ln{f} d^3x d^3v$ and look for functions that extremize its value under the constraint that the total energy $E_{tot} = (1/3)\int E f d^3x d^3v$, the total mass $M = \int f d^3x d^3v$, and the additional quantity

\begin{equation}
Q = \int J^{\nu} |E|^{-3 \nu/4} f d^3x d^3v
\end{equation}

\noindent are taken to be constant. Here the functions $E$ and $J^2$ represent specific energy and specific angular momentum square of a single star subject to a spherically symmetric mean potential $\Phi(r)$. As shown elsewhere \citep{sti87}, this extremization process leads to the following family of distribution functions

\begin{equation}
f^{(\nu)} = A \exp {\left[- a E - d \left( \frac{J^2}{|E|^{3/2}}\right)^{\nu/2}\right]}~,
\end{equation}

\noindent where $a$, $A$, and $d$ are positive real constants. One
may think of these constants as providing two dimensional scales
(for example, $M$ and $Q$) and one dimensionless parameter; the
dimensionless parameter can be taken to be $\gamma = ad^{2/\nu}/(4
\pi GA)$. In principle, $\nu$ is any positive real number; in
practice, we will focus on values of $\nu \approx 1$. The
$f^{(\nu)}$ non-truncated models are constructed by taking this
form of the distribution function for $E \leq 0$, a vanishing
distribution function for $E>0$, and by integrating the relevant
Poisson equation under the condition that the potential $\Phi$ be
regular at the origin and behaves like $- G M/r$ at large radii.
This integration leads to an eigenvalue problem (see Appendix) for
which a value of $\gamma$ is determined by the choice of the
central dimensionless potential, $\gamma = \gamma (\Psi)$, with
$\Psi = - a \Phi (r=0)$.

The main point of the following analysis is the determination of
the Boltzmann entropy $S(M, Q, \Psi)$ and of the total energy
$E_{tot}(M, Q, \Psi)$ along the sequence of models, i.e. as a
function of the concentration parameter $\Psi$ defined above.
These functions, at constant $M$ and $Q$, are illustrated in Fig.~{\ref{fig1}}.
They have been obtained by noting that, from the definitions of $S$
and $f^{(\nu)}$,

\begin{equation}
S = - M \ln{A} + 3 a E_{tot} + d Q.
\end{equation}

\noindent From the definitions $Q = A a^{-9/4}d^{-1 - 3/\nu} \hat{Q}(\Psi)$
and $M = A a^{-9/4}d^{- 3/\nu} \hat{M}(\Psi)$ and the definition of $\gamma$,
we can express the variables $(A, a, d)$ in terms of the variables $(M, Q, \Psi)$
and thus find that the entropy per unit mass can be written as $S/M=S_0(M,Q)+\sigma(\Psi)$, where $S_0$ is constant when the values of $M$ and $Q$
are fixed, with

\begin{equation}
\sigma = -\ln{\left(\hat{M}^{\frac{4\nu - 6}{5 \nu}}\hat{Q}^{\frac{6}{5 \nu}}\gamma^{-\frac{9}{5}}\right)} + \frac{3 \hat{E}}{\hat{M}} + \frac{\hat{Q}}{\hat{M}}~.
\end{equation}

\noindent Here $\hat{E} = \hat{E}(\Psi)$ is the dimensionless total energy defined from $E_{tot} = A a^{-13/4} d^{-3/\nu}\hat{E}$. From the identity $a E_{tot}/M = \hat{E}/\hat{M}$ and the expression of $a = a(M, Q, \Psi)$ obtained previously, we find $E_{tot}/M = H(M,Q)\epsilon(\Psi)$, with:

\begin{equation}
\epsilon = \gamma^{\frac{4}{5}} \hat{M}^{-\frac{9\nu+4}{5\nu}} \hat{Q}^{\frac{4}{5\nu}} \hat{E}~.
\end{equation}

\noindent The factor $H(M,Q)$ is a constant when $M$ and $Q$ are taken to be constant. The quantities $\gamma (\Psi)$, $\hat{M}(\Psi)$, $\hat{Q}(\Psi)$, and $\hat{E}(\Psi)$ that enter the expression of $\sigma$ and $\epsilon$ depend only on $\Psi$ and are evaluated numerically on the equilibrium sequence.

This completes the derivation that allows us to draw the analogy with the classical paper of \citet{lyn68}. This step, straightforward for the $f^{(\nu)}$ models, is by itself interesting and new. In fact, other attempts at applying the paradigm of the gravothermal catastrophe to stellar dynamical equilibrium sequences were either based on an unjustified {\it ansatz} for the identification of the relevant temperature (e.g., see Appendix V in the article by \citealt{lyn68}; \citealt{kat80}; \citealt{mag98}) or on the use of non-standard entropies (for less realistic models; \citealt{cha02}).

\section{The high concentration regime: gravothermal catastrophe}

When the $f^{(\nu)}$ models were constructed \citep{sti87}, it was immediately realized that they have general
properties similar to those of the $f_{\infty}$ models \citep{ber84}; in particular, for values of $\nu \approx 1$,
sufficiently concentrated models along the sequence tend to settle
into a ``stable" overall structure, except for the development of
a more and more compact nucleus, as the value of $\Psi$ increases,
and are characterized by a projected density profile very well
fitted by the $R^{1/4}$ law characteristic of the surface
brightness profile of bright elliptical galaxies. This property is
illustrated in Fig.~{\ref{fig2}}.

Now, by inspection of Fig.~{\ref{fig1}} and by analogy with the study of the isothermal sphere \citep{lyn68}, we can identify the location at $\Psi \approx 9$ as the location for the {\it onset of the gravothermal catastrophe}. This sequence of models thus has the surprising result that the value of $\Psi$ that defines the onset of the gravothermal catastrophe is precisely that around which the models appear to become realistic representations of bright elliptical galaxies. We leave to other papers (see \citealt{ber93}) the detailed discussion of the issues that have to be addressed when comparison is made with the observations.

We note that in this regime of high concentration the general properties of the gravothermal catastrophe are reasonably well recovered by the use of the {\it ansatz} that the temperature parameter conjugate to the total energy is $a$, a quantity directly related to the velocity dispersion in the central regions. Basically, this was the {\it ansatz} made in the discussion of the possible occurrence of the gravothermal catastrophe for the King models or for other sequences of models (e.g., see \citealt{lyn68}, \citealt{kat80}, \citealt{mag98}). Here we have proved that the application of a rigorous derivation, which is available in our case, gives rise to relatively modest quantitative changes in the $(E_{tot},1/T)$ diagram for values of $\Psi$ close to and beyond the onset of the catastrophe (see Fig.~{\ref{fig3}}). However, in Sect.~6 we will draw the attention to an interesting, qualitatively new phenomenon missed in the previous derivations based on the use of the $a$-{\it ansatz}.

In passing, we note that in this regime of relatively high
concentrations, the $f^{(\nu)}$ models possess one intrinsic property
that makes them more appealing than the widely studied $f_{\infty}$
models. This is related to the way the models compare to the phase
space properties of the products of collisionless collapse, as
observed in N-body simulations \citep{van82}. In fact, one noted
unsatisfactory property of the concentrated $f_{\infty}$ models was
their excessive degree of isotropy with respect to the models produced
in the simulations.\footnote{ \citet{mer89} stressed this point and
thus argued that a better representation of N-body simulations would
be obtained by considering the $f_{\infty}$ family of models extended
to the case of negative values of the coefficient $a$ multiplying the
energy in the exponent. Unfortunately, their proposed solution, in
terms of models characterized by such a peculiar phase-space
structure, is unable to reproduce both the core velocity distribution
observed in numerical experiments and the modest amount of radial
anisotropy revealed by observed line profiles. In addition, their
proposed solution is not viable because for negative values of $a$ the
radial anisotropy level is so high that the models are violently
unstable, on an extremely short timescale, with the result that their
structural properties would be drastically changed by rapid evolution
\citep{sti91,ber94}.} Here we can easily check that the anisotropy
level of the concentrated $f^{(\nu)}$ models, while still within the
desired (radial orbit) stability boundary and still consistent with
the modest amount of radial anisotropy revealed by the observations,
seems much closer to that resulting from N-body simulations of
collisionless collapse; in particular, the anisotropy radius
$r_{\alpha}$, defined from the relation $\alpha (r_{\alpha}) = 1$,
with $\alpha = 2 - (\langle v^2_{\theta} \rangle + \langle v^2_{\phi}
\rangle )/ \langle v^2_r \rangle$, is close to the half-mass radius
$r_M$ (while for the $f_{\infty}$ models it is about three times as
large). This is illustrated in Fig.~{\ref{fig4}}. In any case, we
would like to emphasize that, while we have been clearly taking
inspiration from simulations of collisionless collapse, our main
interest is in comparing the structure of our models with that of
observed objects rather than in providing a detailed fit to the
results of N-body simulations.

\section{The intermediate concentration regime: the $R^{1/4}$ law and deviations from it}

The intermediate concentration regime (the precise point that
marks the low concentration regime will be identified in the next
Section) is a regime where the models appear to be stable, with
respect not only to the gravothermal catastrophe (following the
arguments provided earlier; but we should recall that the
catastrophe is expected to require a sufficiently high level of
effective collisionality in order to take off) but also to other
instabilities (see the discussion given by \citealt{ber94} and references therein). The relatively wide variations, between $\Psi = 3.5$ and
$\Psi = 9$, in all the representative quantities that characterize
the equilibrium models suggest that this part of the sequence
could be used to model the weak homology of bright elliptical
galaxies (see \citealt{ber02}), much like the sequence of King
models is able to capture observed systematic variations in the
structure of globular clusters (see \citealt{djo94}).

\section{The low concentration regime: negative global temperature and radial orbit instability}

The low concentration regime is marked by an unexpected and significant difference with respect to the low concentration limit of the classical isothermal sphere \citep{bon56,lyn68}. In fact, while the classical case reduces to the ideal non-gravitating gas, to which Boyle's law applies, for the $f^{(\nu)}$ models the system remains self-gravitating, although it develops a wide core in the density distribution. A clear-cut proof of this difference is given by inspecting the behavior of the global temperature $T$, identified from the thermodynamical definition $T = 1/(\partial{S}/\partial{E_{tot}})$. While the temperature defined by the {\it a-ansatz} remains obviously positive definite, by definition, if we look at Fig.~{\ref{fig1}} we see that the global temperature $T$ changes sign at $\Psi \approx 3.5$. This marks a drastic qualitative deviation from the classical studies.

Here we note a curious coincidence of this transition value of $\Psi$ with the value around which the sequence is bound to change its stability properties with respect to the radial orbit instability \citep{pol81}. Indeed, the location where the sequence is expected to become unstable in this regard is precisely that defined by $\Psi \approx 3.5$, as can be judged from inspection of Fig.~{\ref{fig4}}; around those values of $\Psi$ the level of radial anisotropy, as measured by $2 K_r/K_T$ reaches the threshold value of $1.8 - 2$, known to be sufficient for the excitation of the instability (the precise value of $2K_r/K_T$ corresponding to marginal stability is model dependent; for some sequences the reported value is below the range suggested by \citealt{pol81}).

These clues appear to be interesting and important, but more work is required before a final claim can be made that there is indeed a direct relation between the dynamical radial orbit instability and the fact that the system possesses a negative global temperature, as we found based on the simple work presented here.

\section{Conclusions}

A relatively straightforward and simple thermodynamical description of an equilibrium sequence constructed earlier and known to possess realistic characteristics with respect to bright elliptical galaxies shows that, on the one side, the paradigm of the gravothermal catastrophe may be adequate to explain the occurrence of realistic properties in models of collisionless stellar systems and, on the other side, a long-known dynamical instability might turn out to be interpreted in terms of a thermodynamical argument.

Probably the main open question regarding the models discussed in this paper, partially addressed in previous studies (see \citealt{sti87}), is to what extent the quantity $Q$ is actually reasonably well conserved during violent (partial) relaxation. It is likely that a thorough investigation of this issue may give indications that the quantity is best conserved only in certain ranges of $\nu$. Studies of this type, combined with other dynamical and thermodynamical considerations, may turn out to lead to the identification of a family of equilibrium models with optimal behavior with respect to statistical mechanics, with respect to what we know about collisionless collapse, and with respect to the problem of providing a realistic representation of bright elliptical galaxies.

\acknowledgments

We would like to thank L. Ciotti, M. Stiavelli, and T. van Albada for interesting discussions and suggestions. This work has been partially supported by MIUR of Italy.

\appendix
\section{Numerical integrations}
\begin{figure}
\plotone{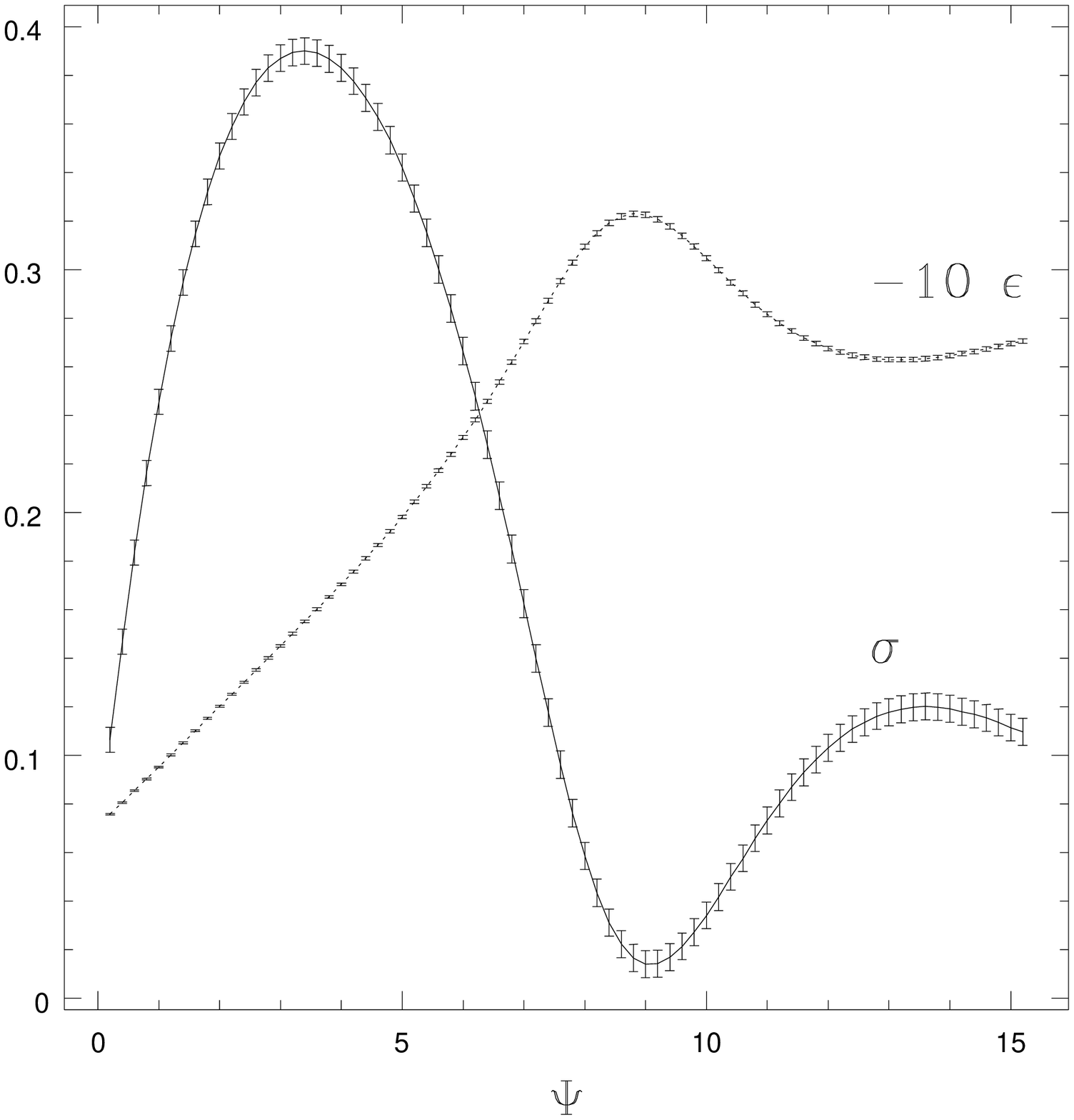}
\caption{Specific entropy and total energy along the equilibrium sequence of $f^{(\nu)}$ models with $\nu=1$ (as a function of the concentration parameter $\Psi$, at constant $M$ and $Q$, and thus expressed by means of the functions $\sigma(\Psi)$ and $\epsilon(\Psi)$ defined in the text). Note that for $\Psi \lesssim 3.5$ the models are characterized by a negative temperature, because the derivatives of $S$ and $E_{tot}$ have opposite signs.\label{fig1}}
\end{figure}

\begin{figure}
\plotone{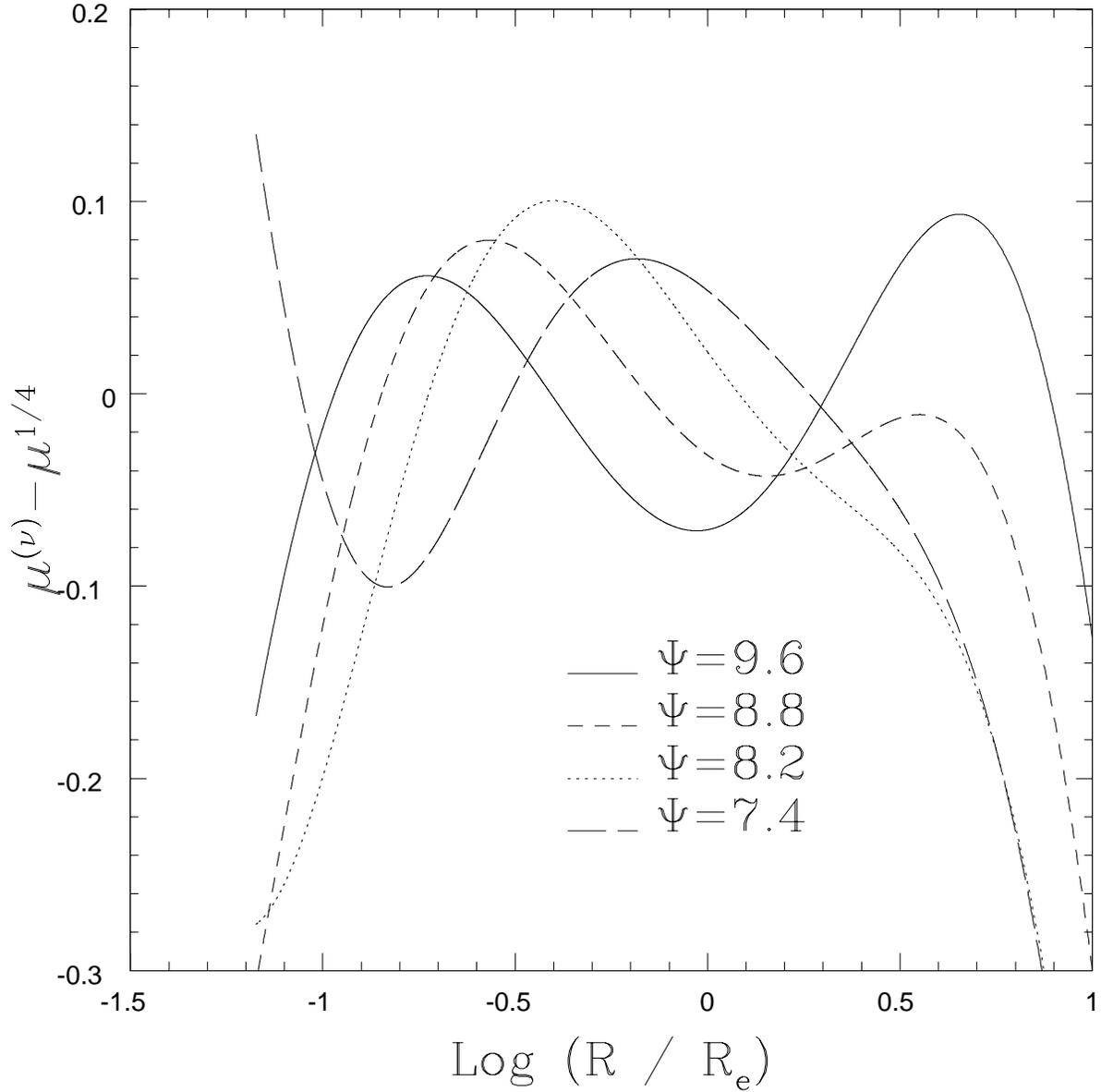}
\caption{Residuals  $\mu^{(\nu)}-\mu^{1/4}$ obtained by fitting the $R^{1/4}$ law to the projected density profile of $f^{(\nu)} $ models for $\nu=1$ and some values of $\Psi$.\label{fig2}}
\end{figure}

\begin{figure}
\plotone{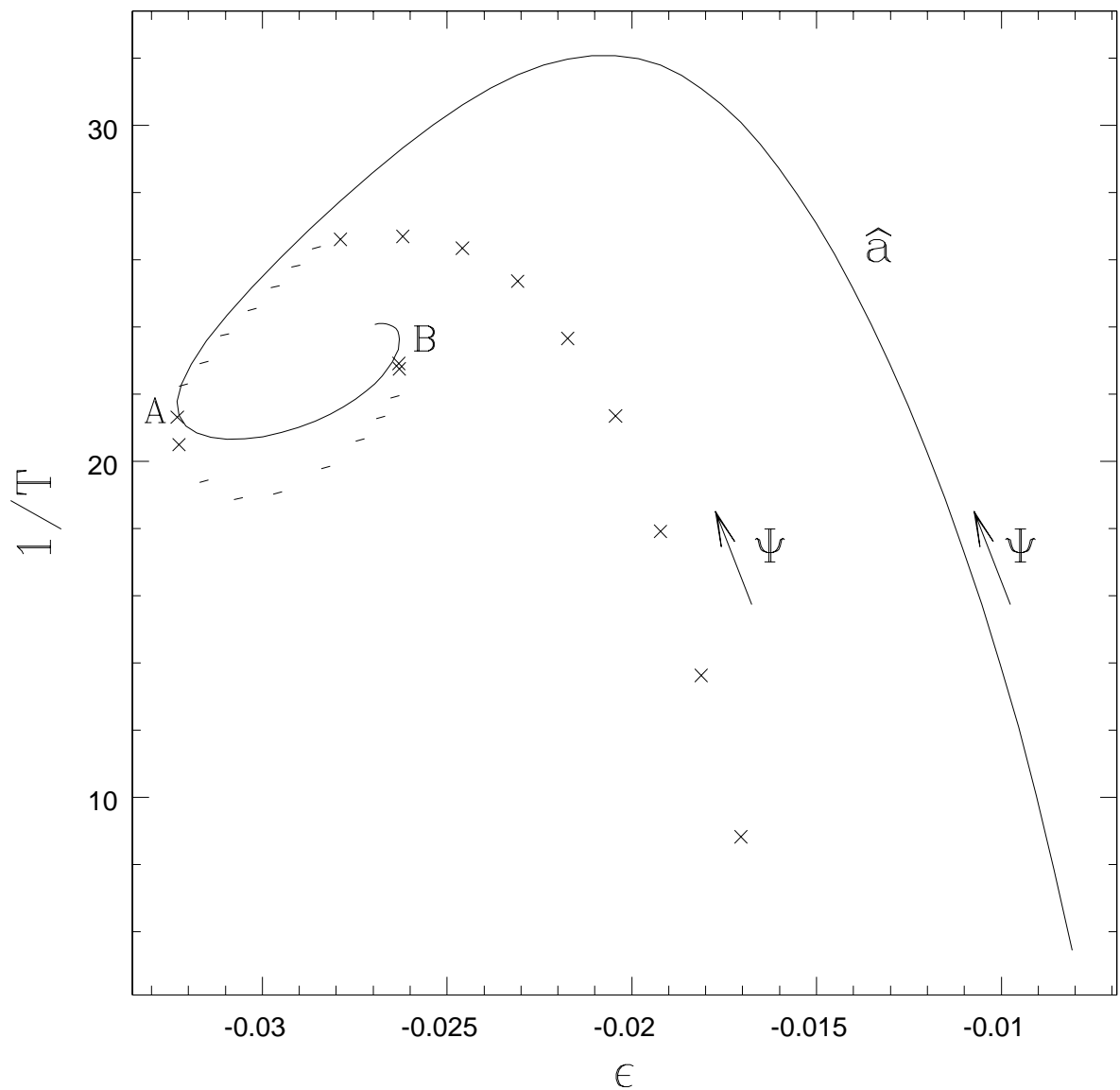}
\caption{Instability spiral of $f^{(\nu)}$ models with
$\nu=1$. The solid line refers to the results obtained with the
$a$-{\it ansatz} (with $\hat{a}= \gamma^{-4/5} \hat{M}^{(\nu+1)/(5
\nu)}\hat{Q}^{-4/(5 \nu)}$). Crosses represent the
global temperature from the definition $\partial S/
\partial E_{tot} $; other symbols indicate estimated points for which the adopted numerical differentiation is less reliable. The values of $\Psi$ and $\epsilon$ for points A and B with a vertical tangent remain unchanged.\label{fig3}}
\end{figure}

\begin{figure}
\plotone{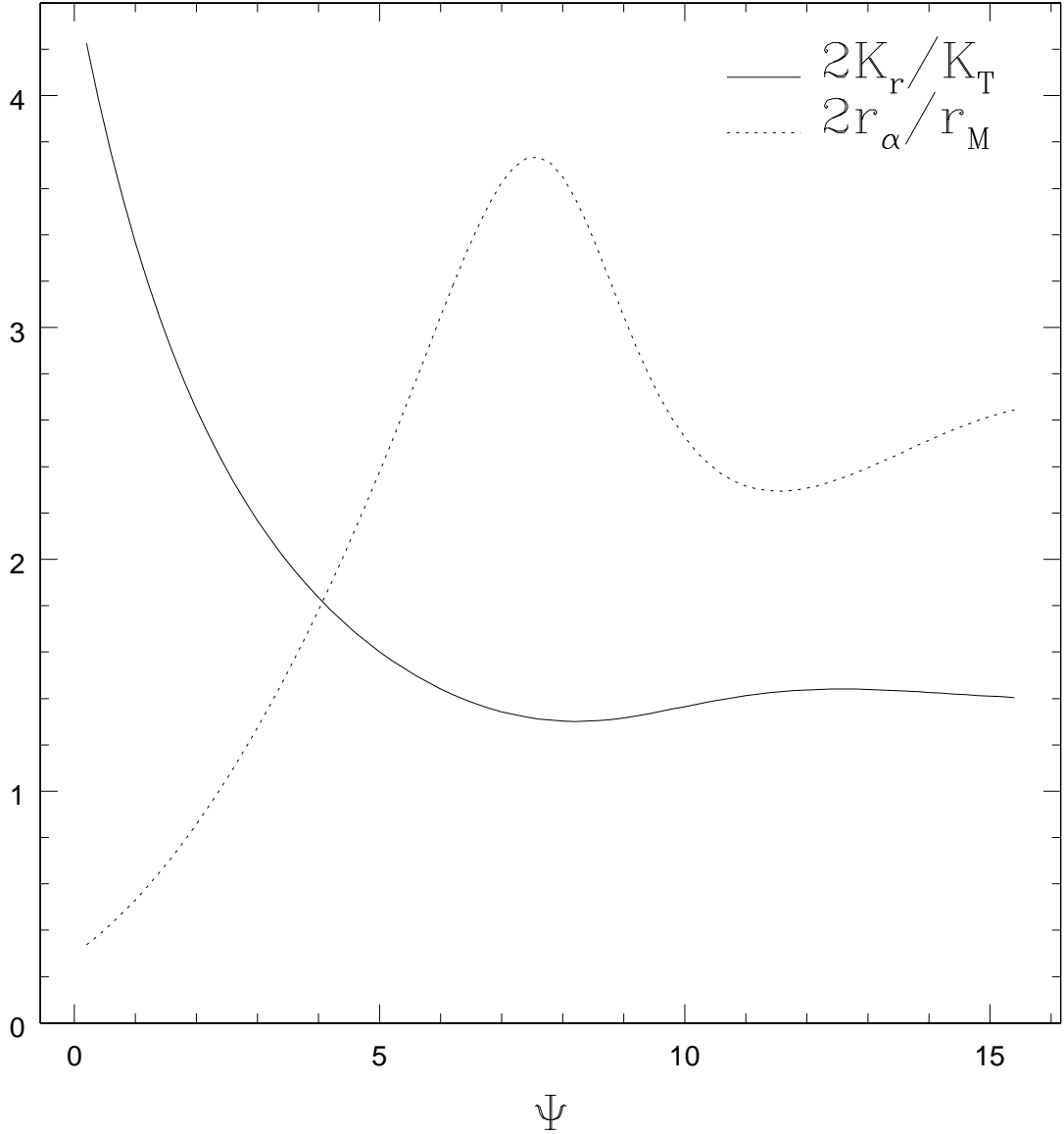}
\caption{Anisotropy along the equilibrium sequence: anisotropy radius in units of the half mass radius $2 r_{\alpha}/r_M$ and the anisotropy parameter  $2K_r/K_T$ (ratio of total kinetic energy in the radial direction to that in the tangential directions) of $f^{(\nu)}$ models with $\nu=1$. \label{fig4}}

\end{figure}

\paragraph{}
The $f^{(\nu)}$ models are constructed by solving the Poisson equation
\begin{equation} \label{eq:PoissonADIM}
\frac{1}{\hat{r}^2} \frac{d}{d \hat r} \hat{r} ^2 \frac{d}{d \hat r} \hat \Phi(\hat r) = \frac{1}{\gamma} \hat{\rho} (\hat{r},\hat{\Phi}),
\end{equation}

\noindent
in which $\gamma$ is considered as an eigenvalue to be determined by imposing the two natural boundary conditions
$ \hat{\Phi}(0) = -\Psi $ and
$ \hat{\Phi}(\hat{r}) \sim - {\hat{M}(\hat{r})}/(4 \pi \gamma \hat{r}) $ as $\hat{r} \rightarrow \infty$.
Here the symbol $\hat{}$  indicates that the quantity is suitably expressed in dimensionless form.
\paragraph{}
We have computed the two-dimensional integral for the density with an adaptive seven-point scheme \citep{bern91} in order to properly handle the presence of a peaked integrand for certain values of the pair $(\hat{r},\hat{\Phi})$. The Poisson equation has then been solved with a fourth order Runge-Kutta code by starting from $\hat{r}=0$ with a seed value for $\gamma$ and iterating the procedure until the boundary condition at large radii is matched within a certain accuracy. Finally, we have proceeded to calculate the global quantities $\hat{M}(\Psi)$, $\hat{Q}(\Psi)$, and $\hat{E}(\Psi)$ from their definitions.
\paragraph{}
In order to check the accuracy of the numerical integration we have performed the following tests:
(1) The virial theorem is satisfied with accuracy of the order $10^{-6}$ or better;
(2) The integrated mass (from its definition) and the mass derived from the asymptotic behaviour of the potential at large radii are the same with accuracy of the order $10^{-4}$;
(3) The expression for $\hat{\Phi}(\hat{r})$ at large radii to two significant orders in the relevant asymptotic expansion has been checked to be correct with an accuracy from $10^{-3}$ to $10^{-4}$;
(4) The asymptotic analysis allows us to estimate the contributions to $\hat{M}$, $\hat{Q}$, and $\hat{E}$  external to a sphere of large radius $R$; this has been checked to help improve the numerical determination of the relevant global quantities.
\paragraph{}
We estimate that the final relative error in the quantities along the equilibrium sequence is of the order of some parts times $10^{-4}$ for $\hat M$ and $\hat Q$ and some parts times $10^{-5}$ for $\hat E$. The total energy is less sensitive to the finite radius truncation error, due to its $1/R^{2}$ convergence. The propagation of these errors leads to the error bars plotted in Fig.~{\ref{fig1}}, where the entropy $\sigma$ is the more difficult to determine with good accuracy.

\clearpage

\clearpage

\clearpage

\clearpage

\end{document}